
\input harvmac
\def\eq{\eqalign}
\def\jour#1*#2*#3*#4{\ \sl#1\ \bf#2\ \rm(19#3)\ #4}
\def\vev#1{\langle#1\rangle}
\def\G{\nabla}
\def\p{\partial}
\def\D{\Delta}
\def\O{\Omega}
\def\ca{\cosh\alpha}
\def\cg{\cosh\gamma}
\def\sa{\sinh\alpha}
\def\sg{\sinh\gamma}
\def\v#1{\vec#1}
\def\P{\Phi}
\def\F#1#2#3{F^{(#1)}_{#2#3}}
\def\A#1#2{A^{(#1)}_{#2}}
\def\CM{{\cal M}}
\def\CL{{\cal L}}

\lref\sh{S. W. Hawking,\jour Comm. Math. Phys.*15*75*1375.}
\lref\ashoke{A. Sen,~hep-th/9411187.}
\lref\ashoketwo{A. Sen,~hep-th/9504147.}
\lref\russo{J. G. Russo and L. Susskind, ~hep-th/9405117}

\Title{\vbox{\baselineskip12pt\hbox{SINP/TNP/95-14}\hbox{hep-th/9509090}}}{}
$$\titlefont\matrix{{\rm Entropy~ of~ Extremal~ Black~ Hole~ Solutions}\cr
                 {\rm of~ String~ Theory}}$$

\centerline{A. Ghosh\footnote{$^{\dag}$}{amit@saha.ernet.in} and
            P. Mitra\footnote{$^{\ddag}$}{mitra@saha.ernet.in}}

\bigskip\centerline{\it Saha Institute of Nuclear Physics}
\centerline{\it 1/AF Bidhannagar, Calcutta, 700 064, INDIA}

\bigskip\centerline{\bf Abstract}\bigskip

For extremal black holes, the thermodynamic entropy is not proportional
to the area. The general form allowed by thermodynamics is worked out
for three classes of extremal black hole solutions
of string theory and shown to be consistent with
the entropy calculated from the density of elementary string states.
On the other hand, the entanglement entropy does not in general agree
with these results.

\Date{9/95}

\newsec{Introduction}
Black hole thermodynamics has been an intriguing subject for many years. The
laws of classical black hole physics suggested definitions of temperature and
entropy purely by analogy with the laws of thermodynamics, but the scale of
these quantities could not be determined that way
\ref\Bek{J. Bekenstein, \jour Phys. Rev.* D9*74* 3292}. It was only with the
introduction of quantum theoretical, or more precisely semiclassical, ideas
that the scale could be set in terms of Planck's constant \sh. The
temperature defined in this way, related to surface gravity, was
later rederived in a euclidean approach where there is a requirement
of periodicity on the euclidean time coordinate if conical singularities are
to be avoided.

Apart from the obvious question about the origin of a nonzero entropy in this
context, the expression for the entropy has itself been a cause for wonder. For
ordinary, or what are now called non-extremal black holes, the entropy is
proportional to the area of the horizon. Explanations have been sought to be
given for this dependence. For instance, it has been pointed out
\ref\thooft{G. 't Hooft, \jour Nucl. Phys. *B256*85*727}\ that the
`entanglement entropy' for matter outside the black hole is proportional to the
area of the horizon. Whether this constitutes an explanation is of
course open to debate.

There has been a lot of interest lately in the special case of extremal black
holes
\ref\GMone{A. Ghosh and P. Mitra, \jour Phys. Rev. Lett.*73*94*2521},
\ref\HHR{S. Hawking, G. Horowitz and S. F. Ross, \jour Phys. Rev.*
D51*95*4302},
\ref\GMtwo{A. Ghosh and P. Mitra, hep-th/9411128, Phys. Lett. B (to appear)},
\ref\GMthree{A. Ghosh and P. Mitra, gr-qc/9507032}.
The temperature and the entropy  behave differently from the
case of nonextremal black holes. Thus, when the temperature defined through
the surface gravity is zero or infinity, it is found that there is no
conical singularity, so that the temperature may really be arbitrary.
Again, the thermodynamical entropy fails to be proportional to the
area of the horizon.
Even the entanglement entropy is not proportional to the
area, and further, the two kinds of entropy behave differently.

Another direction which recent research has taken involves black hole
solutions of string theory. It has been argued that massive string states may
be identified with extremal black hole solutions
\ref\duff{M. Duff, R. R. Khuri, R. Minasian
and J. Rahmfeld, \jour Nucl.Phys.*B418*94*195; M. Duff and J. Rahmfeld,
hep-th/9406105},
\russo, \ashoketwo. This presents an
opportunity of reaching a better understanding of the  entropy of
black holes in terms  of the underlying theory.
The entropy has indeed been calculated \ashoketwo\ from the density of
string states. The result is sometimes zero and sometimes
nonzero even when the area of the horizon vanishes.
By stretching the imagination an area interpretation  can be
developed for the nonzero entropy, but it appears to fail in the case
of the zero entropy. In other words, different approaches have to be used
in the two situations.

As mentioned earlier, the thermodynamic entropy of extremal black holes need
not be proportional to the area. Instead of seeking an area interpretation,
a comparison of the string result with the correct thermodynamical formula
should be made. That is what we do for the entropy of some extremal black
hole solutions of string theory in this paper. The calculation on the basis
of the string level density is now standard.
We shall demonstrate that the expression proportional to the
mass that we have advocated earlier \GMtwo (see also
\ref\puri{P. Mitra, gr-qc/9503042, to appear in Proceedings of the Puri
Workshop on {\sl Physics at the Planck scale}})
for the thermodynamic entropy
fits very well. We also show that in general the entanglement entropy does not
agree with the result.

The extremal black hole solutions considered in this paper are
taken from \ashoke\ and are reviewed in Sec.2, where the expressions for
the entropy as given by the density of
string states \russo\ are also presented ({\it cf}\ \ashoketwo). The form of
the thermodynamic entropy is derived in Sec. 3. Sec. 4 is devoted to the
entropy of scalar matter in the background of these black holes.

\newsec{String based black holes}
\subsec{The solutions}
In four dimensions the massless bosonic fields of heterotic string obtained
by toroidal compactification lead to an effective action with an unbroken
$U(1)^{28}$ gauge symmetry \ashoke:
\eqn\one{\eq{S={1\over 32\pi}\int d^4x&\sqrt{-G}e^{-\P}[R_G+G^{\mu\nu}\G_{\mu}
\P\G_{\nu}\P+{1\over 8}G^{\mu\nu}Tr(\p_{\mu}\CM\CL\p_{\nu}\CM\CL)\cr
&-G^{\mu\mu'}G^{\nu\nu'}{\F a\mu\nu}(\CL\CM\CL)_{ab}{\F b{\mu'}{\nu'}
-{1\over 12}G^{\mu\mu'}G^{\nu\nu'}G^{\rho\rho'}H_{\mu\nu\rho}H_{\mu'\nu'
\rho'}}].}}
Here,
\eqn\two{\CL=\left(\matrix{-I_{22}& \cr &I_6\cr}\right),}
with $I$ representing an identity matrix,
$\CM$ a symmetric 28 dimensional matrix of scalar fields satisfying
\eqn\twoa{\quad \CM\CL\CM=\CL,}
and there are 28 gauge field tensors
\eqn\three{{\F a\mu\nu}=\p_{\mu}{\A a\nu}-\p_{\mu}{\A a\mu},~~ a=1,...28}
and a third rank tensor $H$
\eqn\threea{H_{\mu\nu\rho}=\p_\mu B_{\nu\rho}+2{\A a\mu}\CL_{ab}{\F b\nu\rho}
+{\rm cyclic~permutations~of~}\mu,\nu,\rho}
corresponding to an antisymmetric tensor field $B$.
The canonical metric defined by
\eqn\four{g_{\mu\nu}=e^{-\P}G_{\mu\nu}}
possesses black hole solutions. We shall
study some extremal solutions given  in \ashoke\ .

We choose the scale and asymptotic forms of various backgrounds as
in \ashoketwo\ where the gravitational constant is equal to 2:
\eqn\fiv{\vev{g_{\mu\nu}}=\eta_{\mu\nu},\ \vev{e^{-\P}}={1\over g^2},\
\vev \CM=I_{28},\ \vev{B_{\mu\nu}}=0,\ \vev{\A a {\mu}}=0.}
Here $g$ refers to the string coupling constant.

The dilaton field is nontrivial, though $H$ still vanishes in the solutions
we consider now. The metric $g_{\mu\nu}$ and the dilaton $\P$ are given by
\eqn\five{\eq{ds^2&\equiv g_{\mu\nu}dx^{\mu}dx^{\nu}\cr
&=-{r^2-2mr\over\D^{1/2}}dt^2+{\D^{1/2}\over r^2-2mr}dr^2+\D^{1/2}d\O^2_{II}}}
with
\eqn\six{\eq{\D&=r^2\big[r^2+2mr(\ca\cg-1)+m^2(\ca-\cg)^2\big],\cr{\rm and}
\qquad e^{\P}&={g^2r^2\over\D^{1/2}}}.}
Here $\alpha, \gamma$ are real parameters. The time components of the
gauge fields are given by
\eqn\seven{\v A_t=\cases{-{g\v n_L\over\sqrt 2}{mr\sa\over\D}[r^2\cg+mr(\ca-
\cg)]&$L=1,...   22$\cr -{g\v n_R\over\sqrt 2}{mr\sg\over\D}[r^2\ca+mr
(\ca-\cg)]&$R=23,...   28$}}
with $\v n_L, \v n_R$ denoting respectively 22-component and 6-component unit
vectors and
\eqn\eight{\CM=I_{28}+\left(\matrix{Pn_Ln_L^T&Qn_Ln_R^T\cr Qn_Rn_L^T&Pn_Rn_R^T
\cr}\right),}
where
\eqn\nine{\eq{P&={2m^2r^2\over\D}\sinh^2\alpha\sinh^2\gamma\cr
Q&=-{2mr\over\D}\sa\sg[r^2+mr(\ca\cg-1)].}}
All other backgrounds vanish for this solution.

The ADM mass of the black hole and its charges are given by
\eqn\ten{\eq{M&={1\over 4}m(1+\ca\cg)\cr\v Q&=\cases{{g\v n_L\over\sqrt 2}
m\sa\cg&$L=1, ...  22$\cr{g\v n_R\over\sqrt 2}m\sg\ca&$R=23, ...  28$}\cr}}
The area of the horizon, which is at $r=2m$, is
\eqn\eleven{A_H=8\pi m^2(\ca+\cg),}
and the inverse temperature (as defined in terms of the surface gravity)
is given by
\eqn\elevena{\beta_H=4\pi m(\ca+\cg).}
Three specific extremal limits were considered in \ashoke.

i) Here,
\eqn\eighteen{m\to 0,\quad\alpha=\gamma\to\infty,
\quad {\rm with}\quad m\cosh^2\alpha=m_0.}
Then
\eqn\twentya{\quad A_H=0,\quad T_H=\infty,}
and
\eqn\nineteen{M={m_0\over 4},\quad\v Q_L={gm_0\over\sqrt 2}\v n_L,\quad\v Q_R=
{gm_0\over\sqrt 2}\v n_R.}
Consequently,
\eqn\twenty{M^2={1\over 8g^2}\v Q_L^2={1\over 8g^2}\v Q_R^2.}
Thus the Bogomol'nyi bound is saturated.
Although the temperature appears to be infinite here from the surface gravity
formula, there is no  conical singularity \GMthree\
at the horizon in the corresponding Euclidean metric, so that the temperature
should really be taken to be arbitrary.

ii) Here,
\eqn\fifteen{m\to 0,\quad\gamma\to\infty,
\quad {\rm with}\quad m\cg=m_0,\quad\alpha={\rm finite}.}
Then
\eqn\seventeena{\quad A_H=0,\quad T_H={1\over 4\pi m_0},}
and
\eqn\sixteen{M={m_0\over 4}\ca,\quad\v Q_L={gm_0\over\sqrt 2}\sa\ \v n_L,\quad
\v Q_R={gm_0\over\sqrt 2}\ca\ \v n_R.}
Consequently,
\eqn\seventeen{M^2={1\over 8g^2}\v Q_R^2.}
So in this case the Bogomol'nyi bound is saturated.

iii) Here,
\eqn\twelve{m\to 0,\quad\alpha\to\infty,
\quad {\rm with}\quad m\ca=m_0,\quad\gamma={\rm finite}.}
Then
\eqn\thirteena{\quad A_H=0,\quad T_H={1\over 4\pi m_0}}
and
\eqn\thirteen{M={m_0\over 4}\cg,\quad\v Q_L={gm_0\over\sqrt 2}\cg\ \v
n_L,\quad\v Q_R={gm_0\over\sqrt 2}\sg\ \v n_R.}
Consequently,
\eqn\fourteen{M^2={1\over 8g^2}\v Q_L^2.}
Note that the Bogomol'nyi bound is {\it not} saturated here:
\eqn\fourteen{M^2\ne{1\over 8g^2}\v Q_R^2.}

\subsec{Entropy from string level density}
The density of states in heterotic string theory is given
for a large number $N$ of oscillators by \russo\ as
\eqn\tone{\rho\approx{\rm const.}N^{-23/2}e^{2a\sqrt N},}
where $a_L=2\pi, a_R=\sqrt{2}\pi$.
The numbers of oscillators in the left and right sectors are related to the
mass and charges of
the corresponding states by the usual formula
\eqn\ttwo{M^2= {g^2\over 8}({\v Q_L^2\over g^4}+2N_L-2)
= {g^2\over 8}({\v Q_R^2\over g^4}+2N_R-1).}
To find the level density in terms of the ADM mass of a black hole, one has
to combine
this formula with the relation between the mass and the charges as applicable
for the solution describing that black hole. The three cases have to be
discussed separately.

i) In this case, $N_L=1$ and $N_R={1\over 2}$, so the entropy is zero.

ii) In the second case, $N_R={1\over 2}$ and the entropy arises from large
values of $N_L$.
\eqn\ttree{S=\log\rho\approx 4\pi\sqrt{N_L}\approx{8\pi\over g}
\sqrt{M^2 -{Q_L^2\over 8g^2}}={8\pi\over g\cosh\alpha}M=
{2\pi\over g}m_0.}

iii) In this case, $N_L=1$ and the entropy
arises from large values of $N_R$.
\eqn\tthree{S=\log\rho\approx
2\sqrt{2}\pi\sqrt{N_R}\approx{4\sqrt{2}\pi\over g} \sqrt{M^2
-{Q_R^2\over 8g^2}}={4\sqrt{2}\pi\over g\cosh\gamma}M=
{\sqrt{2}\pi\over g}m_0.}

Thus we see that in all three extremal cases the entropy has a linear
dependence on the mass of the black hole, though in one of the two
Bogomol'nyi saturated
cases the entropy is actually zero.
To understand these values, one has to recall our
formula \GMtwo\
$S=kM$ for extremal Reissner - Nordstrom black holes
with $k$ a constant. A generalization of that result to the case of several
charges will
be presented in the next section.

\newsec{Thermodynamic entropy}
For nonextremal black holes, the laws of black hole physics suggest that
the entropy is proportional to the area of the horizon. When the scale
is fixed by comparing the
temperature thus suggested with that given by the semiclassical calculations
of \sh,
the entropy turns out to be a quarter of the area. If one is interested
in an extremal black hole, one may be tempted to regard it as a special
limiting case of a sequence
of nonextremal black holes and thus infer that the same formula should
hold for the entropy. It was pointed out in the context of Reissner -
Nordstrom black holes \HHR\
that the extremal and nonextremal cases of the euclidean version are
topologically different, so that continuity need not hold. It was also
argued that the entropy in
the extremal case should vanish. Subsequently it was shown \GMtwo, \puri\ that
if the derivation of an expression for the thermodynamic entropy is
attempted afresh for the
extremal case, one obtains a result proportional to the mass of the
black hole with an undetermined scale. We shall now see
how the arguments of \GMtwo\ can be adapted to the stringy black holes. The
three cases have to be treated separately.

i) This is the simplest case. The first law of thermodynamics takes the form
\eqn\dreia{\tilde T dS=dM-\v\P\cdot d\v Q,}
where $\v\P$ represents the chemical potential corresponding to the charge
$\v Q$ and the temperature has been written as $\tilde T$ to indicate the
possibility of its being different from the infinite temperature $T_H$
\GMthree.
We can make use of the
$O(22)\times O(6)$ symmetry to write
\eqn\dreid{\v\P=\cases{{\sqrt 2f_L\v n_L \over 4g}&$L=1,...22$\cr
{\sqrt 2f_R\v n_R \over 4g}&$R=23,...28$},}
where $f_L,f_R$ are unknown functions of $m_0$.
There are standard expressions for the chemical potential in nonextremal
cases, but we cannot use them for two reasons: first, extremal black holes
are not continuously connected to nonextremal black holes \HHR, and
secondly, the standard expressions are calculated by differentiating the mass
with respect to charges at constant {\it area} in the anticipation that
constant area and constant entropy are synonymous!

By using \nineteen\ and \dreia\ we can now write
\eqn\dreif{\tilde TdS={(1-f_L-f_R)dm_0\over 4}.}
Here it is understood that only such thermodynamic processes are allowed
which leave the black hole in the class being considered, {\it i.e.,}
all variations  are in the parameters $m_0$ and the unit vectors
$\v n_L,\v n_R$.
Now the partition function can be written as
\eqn\dreifa{Z=\exp (-{W\over\tilde T}),}
where the grand canonical thermodynamic potential is
\eqn\dreifb{W=M-\tilde TS-\v\P\cdot\v Q.}
Moreover, in the leading semiclassical approximation, $Z$ can be taken to be
the
exponential of the negative classical action, which vanishes in this case
as the area vanishes.  Hence $W$ vanishes too and
\eqn\dreig{\tilde TS=M-\v\P\cdot\v Q={(1-f_L-f_R)m_0\over 4}.}
Comparison of \dreif\ and \dreig\ yields
\eqn\dreih{{dS\over dm_0}={S\over m_0},}
{\it i.e.,}
\eqn\dreii{S=km_0}
with some undetermined constant $k$. As $k$ may be taken to vanish,
the vanishing string answer is consistent with the thermodynamical
expression for the entropy.

ii) This case is slightly more complicated because of the existence
of an extra parameter $\alpha$. We can still introduce the chemical potential
by \dreid, but the $f$-s are now unknown functions of both $m_0$ and
$\alpha$. Because of the vanishing area and classical action, we have
an analogue of \dreig:
\eqn\zweia{T_HS=M-\v\P\cdot\v Q={(\ca-f_L\sa-f_R\ca)m_0\over 4}.}
Note that we have not written $\tilde T$ here: this is because the
temperature is not arbitrary in this case but has to be $T_H$.
Using \seventeena, we then have
\eqn\zweib{S=\pi m_0^2(\ca-f_L\sa-f_R\ca).}
Further, the first law of thermodynamics yields
\eqn\zweic{T_H{\p S\over\p m_0}={\p M\over\p m_0}-\v\P\cdot{\p\v Q\over\p m_0}
={(\ca-f_L\sa-f_R\ca)\over 4}.}
This can be written in view of \zweib\ as
\eqn\zweid{{\p S\over\p m_0}={S\over m_0},}
whence
\eqn\zweie{S=k(\alpha)m_0,}
with $k(\alpha)$ now an undetermined function of $\alpha$.
This function cannot be fixed by considering the analogue of \zweic\ where
the $m_0$-derivatives are replaced by $\alpha$-derivatives; what happens
is that $f_L, f_R$ get expressed in terms of $k$.
The string answer for the entropy is
indeed of the form \zweie, with $k(\alpha)$ actually taking the constant value
${2\pi\over g}$.

iii) This case is similar to the previous one and clearly leads to
\eqn\einsa{S=k(\gamma)m_0,}
where $k(\gamma)$ is now the constant ${\sqrt 2\pi\over g}$.

\newsec{Entanglement entropy}
The entropy of scalar matter outside a black hole was calculated in
\thooft\ for a Schwarzschild black hole. It can be easily generalized
for other black holes and written as \puri\
\eqn\entang{S={8\pi^3\over 45\beta^3}\int_{r_h+\epsilon}^L dr~g_{rr}^{1/2}
(-g_{tt})^{-3/2} g_{\theta\theta},}
where $r_h$ is the location of the horizon and the integration runs from
the `brick wall', which is a distance $\epsilon$ outside the horizon, to
a large value $L$. These
can be thought of as the ultraviolet and infrared cutoffs respectively.
The $L$-dependent part has to be subtracted because it arises even in the
absence of the black hole.
$\beta$ is the inverse of the temperature, which is taken to be the
Hawking temperature when it is
finite. We shall apply this formula to the three extremal black holes under
consideration.

i) Here, one finds
\eqn\entthree{S={8\pi^3\over 45\beta^3}\int_{\epsilon} dr~(2m_0)^{3/2}
r^{1/2}=-({16\pi^3\over 135\beta^3})(2m_0\epsilon)^{3/2}.}
Unlike the usual situation, this expression remains finite in the limit
of vanishing $\epsilon$, {\it i.e.,} there is no ultraviolet divergence.
The entire entropy can be
absorbed in the long distance part, so it is natural to take the
entanglement entropy to be {\it zero}. This is consistent with the
string result as well as the form derived for
the thermodynamic entropy. Thus all answers, including
the prediction of \HHR, agree.

ii) Here, the expressions given above lead to
\eqn\entone{S={8\pi^3\over 45\beta^3}\int_{\epsilon} dr~{m_0^3\over r}
={1\over 360}\log{1\over\epsilon}.}
It is customary to replace the cutoff $\epsilon$ by the proper
distance of the brick wall at $r_h+\epsilon$ from the horizon, {\it i.e.,}
\eqn\enttwo{\tilde\epsilon=\int_0^\epsilon dr~\sqrt{g_{rr}}\approx
2\sqrt{m_0\epsilon}.}
This means
\eqn\entoneb{S={1\over 360}\log{1\over\tilde\epsilon^2}.}
This logarithmic dependence is known from \GMone\ and is reminiscent of $(1+1)$
dimensional black holes. It does not agree with the expression given by
the string level density.

iii) Here too the same expression follows.

\newsec{Discussion}

Although we demonstrated in \GMtwo and \puri that some extremal black
holes do not satisfy the area formula and have thermodynamic entropies
proportional to the mass, it may not have been clear whether our arguments
apply to {\it stringy} black holes.

We have therefore considered here three classes of extremal stringy black
holes.
Our treatment of the thermodynamics
does lead to expressions for the entropy proportional to the mass. In the
last two cases, the entropy calculated from the density of string states
is indeed proportional to the mass. In the first case, the same approach
leads to a vanishing entropy, which is also consistent with the general
form derived by us on the basis of thermodynamics.

It may appear somewhat disappointing that the thermodynamic approach
gives an expression for the entropy in terms of an
undetermined constant or function $k$. The same thing happened in the case
of the Reissner - Nordstrom black hole \GMtwo. As we argued there, this
is bound to happen in the case of zero or infinite $T_H$ where the actual
temperature is arbitrary and does not introduce a scale as is done for
nonextremal black holes by $T_H$. In the cases considered here with finite
$T_H$, a scale is of course involved. But there are many different ways
of embedding a black hole in string theory \duff\ and
in general the string
level density depends on the embedding. As the thermodynamically
derived expression has to accommodate all these possible values, $k$ has
to remain undetermined without further specification.

As the entropy of matter in the background of a black hole is often studied
in the context of the entropy of a black hole, we have also compared this
kind of entropy with the other kinds. In the last two cases, the answer
is not of the form derived from thermodynamics and hence inconsistent with
the value given by the density of string states. In the first case, the
matter entropy can be taken to be zero, and hence made consistent with
the string result as well as thermodynamics. In general, the matter entropy
must be said to be of a different form, and so the thermodynamical
entropy cannot easily be explained in terms of this kind of entropy.


\listrefs
\bye